\documentclass[12pt, draftclsnofoot, onecolumn]{IEEEtran}

\usepackage{pdfpages}
\usepackage{commath, amsmath, graphics, amssymb, epsfig, float, babel}
\usepackage{amsfonts, graphicx, color}
\usepackage{cite, array, subfigure, multirow, hhline}
\usepackage[noend]{algpseudocode}

\def\BibTeX{{\rm B\kern-.05em{\sc i\kern-.025em b}\kern-.08em
    T\kern-.1667em\lower.7ex\hbox{E}\kern-.125emX}}

\newcommand{\be}{\begin{equation}}

\newcommand{\ee}{\end{equation}}
\newcommand{\bea}{\begin{eqnarray}}
\newcommand{\eea}{\end{eqnarray}}
\newcommand{\bdp}{\begin{displaymath}}
\newcommand{\edp}{\end{displaymath}}

\floatstyle{ruled}
\newfloat{algorithm}{tbp}{loa}
\providecommand{\algorithmname}{Algorithm}
\floatname{algorithm}{\protect\algorithmname}

\makeatletter
\def\hlinewd#1{\noalign{\ifnum0=`}\fi\hrule \@height #1 \futurelet \reserved@a\@xhline}

\begin{document}
\title{\Large{Decentralized Computation Offloading With Cooperative UAVs: \\ Multi-Agent Deep Reinforcement Learning Perspective}}

\author{\IEEEauthorblockN{Sangwon Hwang, Hoon Lee, Juseong Park and Inkyu Lee,~\IEEEmembership{Fellow,~IEEE}} \vspace{-15mm}

\thanks{© 2022 IEEE. Personal use of this material is permitted. Permission from IEEE must be obtained for all other uses, in any
current or future media, including reprinting/republishing this material for advertising or promotional purposes, creating new
collective works, for resale or redistribution to servers or lists, or reuse of any copyrighted component of this work in other
works.

S. Hwang and I. Lee are with the School of Electrical Engineering, Korea University, Seoul 02841, Korea (e-mail: \{tkddnjs3510, inkyu\}@korea.ac.kr). 

H. Lee is with the Department of Information and Communications Engineering, Pukyong National University, Busan 48513, Korea (e-mail: hlee@pknu.ac.kr). 

J. Park is with the University of Texas at Austin, Austin, TX 78712, USA (e-mail: Juseong.park@utexas.edu)
\textit{(Corresponding authors: Hoon Lee; Inkyu Lee.)}
}}
\maketitle

\begin{abstract}
Limited computing resources of internet-of-things (IoT) nodes incur prohibitive latency in processing input data. This triggers new research opportunities toward task offloading systems where edge servers handle intensive computations of IoT devices. Deploying the computing servers at existing base stations may not be sufficient to support IoT nodes operating in a harsh environment. This requests mobile edge servers to be mounted on unmanned aerial vehicles (UAVs) that provide on-demand mobile edge computing (MEC) services. Time-varying offloading demands and mobility of UAVs need a joint design of the optimization variables for all time instances. Therefore, an online decision mechanism is essential for UAV-aided MEC networks. This article presents an overview of recent deep reinforcement learning (DRL) approaches where decisions about UAVs and IoT nodes are taken in an online manner. Specifically, joint optimization over task offloading, resource allocation, and UAV mobility is addressed from the DRL perspective. For the decentralized implementation, a multi-agent DRL method is proposed where multiple intelligent UAVs cooperatively determine their computations and communication policies without central coordination. Numerical results demonstrate that the proposed decentralized learning strategy is superior to existing DRL solutions. The proposed framework sheds light on the viability of the decentralized DRL techniques in designing self-organizing IoT networks.
\end{abstract}

\section{Introduction}
Massive machine-type communication (mMTC) is one of the important features of next-generation networks \cite{ZDawy:WC16}. It facilitates self-organizing connectivity services for massive mobile nodes with various missions such as healthcare applications and disaster monitoring. Such internet-of-things (IoT) nodes are of small form-factor having strict restrictions on battery capacity, which pose an inevitable performance loss both in communications and computations.

For the purpose of control and monitoring infrastructures, IoT nodes are typically deployed in harsh and rural areas that are out of network coverage. To enhance the connectivity performance, mobile access points (APs) mounted on unmanned aerial vehicles (UAVs) are essential requirements of mMTC. With a carefully optimized UAV trajectory, the communication latency can be remarkably reduced by decreasing the wireless access distance \cite{HLee:TVT18,SEom:TVT18}. However, the computation latency still remains unaddressed even with the enhanced networking technologies. The computation capacity of IoT nodes is mostly limited by their low-power hardware circuitry. This blocks timely processing of latency-critical tasks dominating the overall completion time.

These challenges bring up new research paradigms in mobile edge computing (MEC) systems where intensive computations of IoT devices (IDs) are handled by edge networks. Computing servers are deployed at the edge of networks. Consequently, the task execution latency mainly depends on the computing power of edge servers and the quality of wireless access links, but not on the packet transmission strategy.
Edge servers can be realized by UAVs to further improve the system performance. In this UAV-aided MEC system, computational tasks of mobile nodes are offloaded to UAV servers through wireless channels. Hence, intensive computations are successfully addressed at the expense of additional costs for task upload/download steps. This poses challenging situations for joint optimization of communication and computation strategies. UAV servers allow an additional degree of freedom in their time-varying trajectories, and temporal correlations of task arrival patterns resulting from highly fluctuating system dynamics need to be considered. Dispatching UAV servers close to ground nodes reduces both the communication latency as well as the energy consumption \cite{SJeong:TVT18}. However, existing optimization algorithms, which are originally intended for instantaneous optimization problems, lack adaptivity to the time-varying system. This motivates the development of an online decision-making strategy for UAV servers.

\begin{figure*}[hbt!]
\begin{center}
\includegraphics[width=0.8\linewidth]{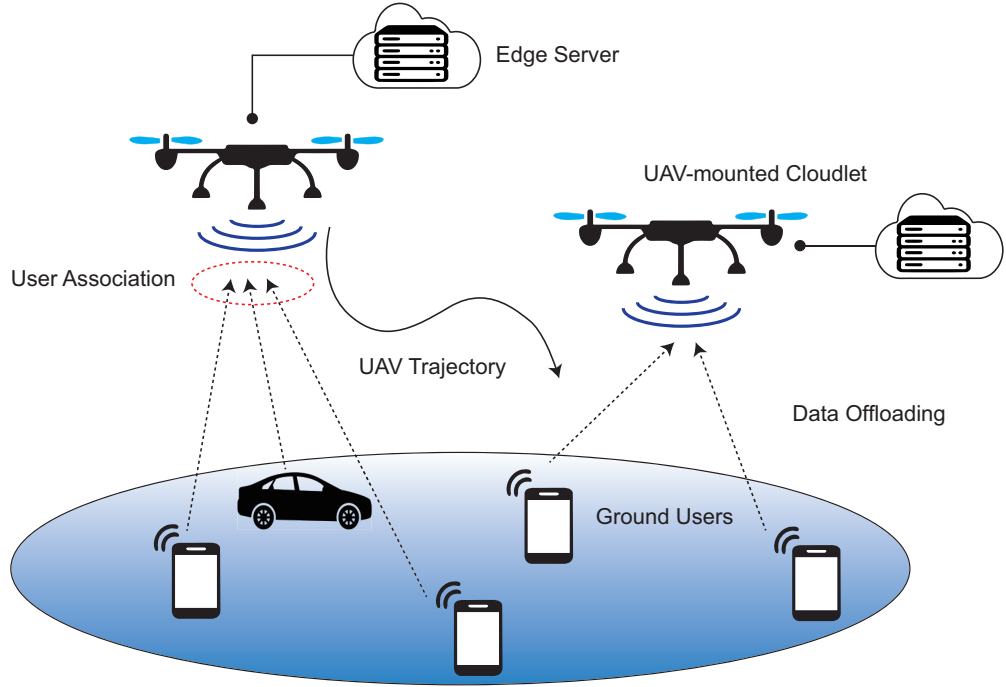}
\end{center}
\caption{A schematic diagram for UAV-aided MEC networks. The computation performance for IDs can potentially be improved by offloading tasks to the UAV-mounted edge computing servers for remote execution.}
\label{figure:system}
\end{figure*}

There have been recent efforts on optimizing online decision-making policies using deep reinforcement learning (DRL) techniques \cite{QLiu:TVT20,LZang:ACC21,LWang:TMC21,LWang:TCCN21,AGao:TVT21,HPeng:JSAC21}. Conventional studies have fundamental limitations on a joint design of heterogeneous features of communication and computation systems. They have been confined to separate optimization policies of coupled variables, thereby leading to suboptimal performance. Moreover, existing multi-UAV MEC systems \cite{LWang:TMC21,LWang:TCCN21,AGao:TVT21,HPeng:JSAC21} assume an ideal scenario where a central coordinator, e.g., base stations and APs, is available for the management of separate UAVs. This requires centralized training processes that rely on global network observations aggregating statistics of individual UAVs and IDs. Therefore, it is essential to develop a new decentralized DRL framework that enables a joint optimization of crucial MEC factors, including UAV movement and resource allocation rules, in dynamic networking scenarios based only on separate UAV processing.

This article presents a multi-agent DRL method that identifies an efficient management policy of task offloading, scheduling, computation, and UAV trajectories. Such a joint optimization results in highly coupled continuous-valued action spaces having numerous solution candidates. To this end, a novel agent structure is proposed, which splits coupled decision-making processes using individual neural layers dedicated to each of action variables.

For realizing decentralized MEC mechanisms, it is critical to design interaction strategies among multiple UAVs. This is formulated as a multi-agent decision process where agents installed at individual UAVs make online decisions. Training of these UAV agents is achieved by a novel decentralized DRL framework. Simulation results demonstrate that the proposed method achieves the performance of centralized methods and is superior to existing DRL solutions.

This article is organized as follows: Optimization challenges of UAV-aided MEC networks are introduced, and it is followed by an overview of state-of-the-art DRL techniques. A decentralized multi-agent learning mechanism is proposed which jointly optimizes resource management and UAV mobility. The viability of the proposed scheme is investigated through numerical results. Concluding remarks and important research directions are presented.

\section{Surveys on UAV-Aided Computation Offloading Schemes} \label{sec:system and problem}

Low-power computing units of IDs has exploded the completion time of task processing stages. This brings a new paradigm towards MEC techniques to offload tasks of large volumes to powerful edge servers. In particular, UAV-mounted mobile servers are necessary to collect big data of IDs. UAVs can change their traveling paths dynamically to decrease access distance, thereby providing timely communication and computing services to proximity devices. This section presents the design challenges of UAV-aided MEC networks and offers an overview of the state-of-the-art optimization approaches.

\begin{table*}[]
\caption{Summary of DDPG Approaches for UAV-Aided MEC Systems}
\label{tab:tab1}
\centering
\resizebox{\textwidth}{!}{
{\huge
\begin{tabular}{|c||c|c|c|c|c|}
\hline
\textbf{Scheme} & \textbf{Objective}                     & \textbf{Constraints}                                                                                                                    & \textbf{Action variables}                                                                                          & \textbf{Training} & \multicolumn{1}{l|}{\textbf{Execution}} \\ \hline
{\cite{LWang:TMC21}}            & Minimize energy consumption        & \begin{tabular}[c]{@{}c@{}}Computation/communication latency\\ Computation capacity\\ Maximum UAV speed\end{tabular}                                  & 3D UAV trajectory                                                                                                     & Centralized                    & Centralized                                       \\ \hline
{\cite{LWang:TCCN21}}            & Maximize energy fairness           & \begin{tabular}[c]{@{}c@{}}Computation/communication latency\\ Maximum UAV speed\\ UAV collision avoidance\end{tabular}                                                         & 2D UAV trajectory                                                                                                     & Centralized                    & Decentralized                                     \\ \hline
\cite{AGao:TVT21}          & Maximize task processing rate            & \begin{tabular}[c]{@{}c@{}}Maximum UAV speed\\ UAV collision avoidance\end{tabular} & \begin{tabular}[c]{@{}c@{}}3D UAV trajectory\end{tabular} & Centralized                  & Decentralized                                     \\ \hline

{\cite{HPeng:JSAC21}}           & Maximize the number of offloaded tasks & Computation/communication latency                                                                                                                   & \begin{tabular}[c]{@{}c@{}}Task offloading\\ Computing/caching resources\end{tabular}                              & Centralized                    & Decentralized                                     \\ \hline

Proposed           & Minimize energy consumption            & \begin{tabular}[c]{@{}c@{}}Computation/communication latency\\ Computation capacity\\ Maximum UAV speed\\ UAV collision avoidance\end{tabular} & \begin{tabular}[c]{@{}c@{}}3D UAV trajectory\\ Device association\\ Computing resource\\ Task offloading\end{tabular} & Decentralized                  & Decentralized                                     \\ \hline
\end{tabular}
}
}
\end{table*}

\subsection{Challenges in UAV-aided MEC Systems}
Figure \ref{figure:system} illustrates UAV-aided MEC networks where multiple UAVs help the computations of IDs. These can be carried out locally or offloaded to UAVs to exploit powerful computing resources. The offloading procedure imposes additional latency related to task upload/download steps. This incurs a fundamental tradeoff between local computation and offloading strategies. The link quality between UAVs and IDs plays a key role in determining the task offloading policy. The communication latency can be reduced by positioning UAVs close to IDs. In contrast, a long-range air-to-ground link fails in successful task offloading and increases both the latency and energy consumption of IDs to execute their tasks locally. Thus, careful plans of UAV trajectories lead to a huge reduction in the task completion time as well as enhanced energy efficiency.

Albeit such advantages, the UAV mobility introduces nontrivial optimization challenges. Due to time-varying channel dynamics, the performance of the MEC systems may fluctuate according to three-dimensional (3D) positioning of UAVs. Also, task arrivals and locations of IDs are not manageable and are treated as non-stationary random processes. Consequently, the design strategies of the UAV mobility, task offloading, and resource management policies are coupled into sequential decision-making processes.

A single UAV is not sufficient to cover a wide range of IoT networks. Thus, employing multiple UAVs is an essential setup. The optimum UAV control requests an external coordination unit, such as terrestrial base stations, for optimizing all variables with the aid of centrally collected information. Such a centralized strategy is, however, infeasible for practical MEC systems. To avoid this drawback, self-organizing and decentralized management of UAV servers, which do not require a central coordinator, is critical for IoT applications. To this end, individual computation rules of separate UAVs can be designed such that they only leverage locally observable states. However, decentralized schemes suffer from the lack of global network information, possibly leading to a performance loss. This fundamentally involves a careful interaction mechanism among UAVs to exchange useful messages relevant to individual decisions. 

\begin{figure*}[hbt!]
\begin{center}
\includegraphics[width=\linewidth]{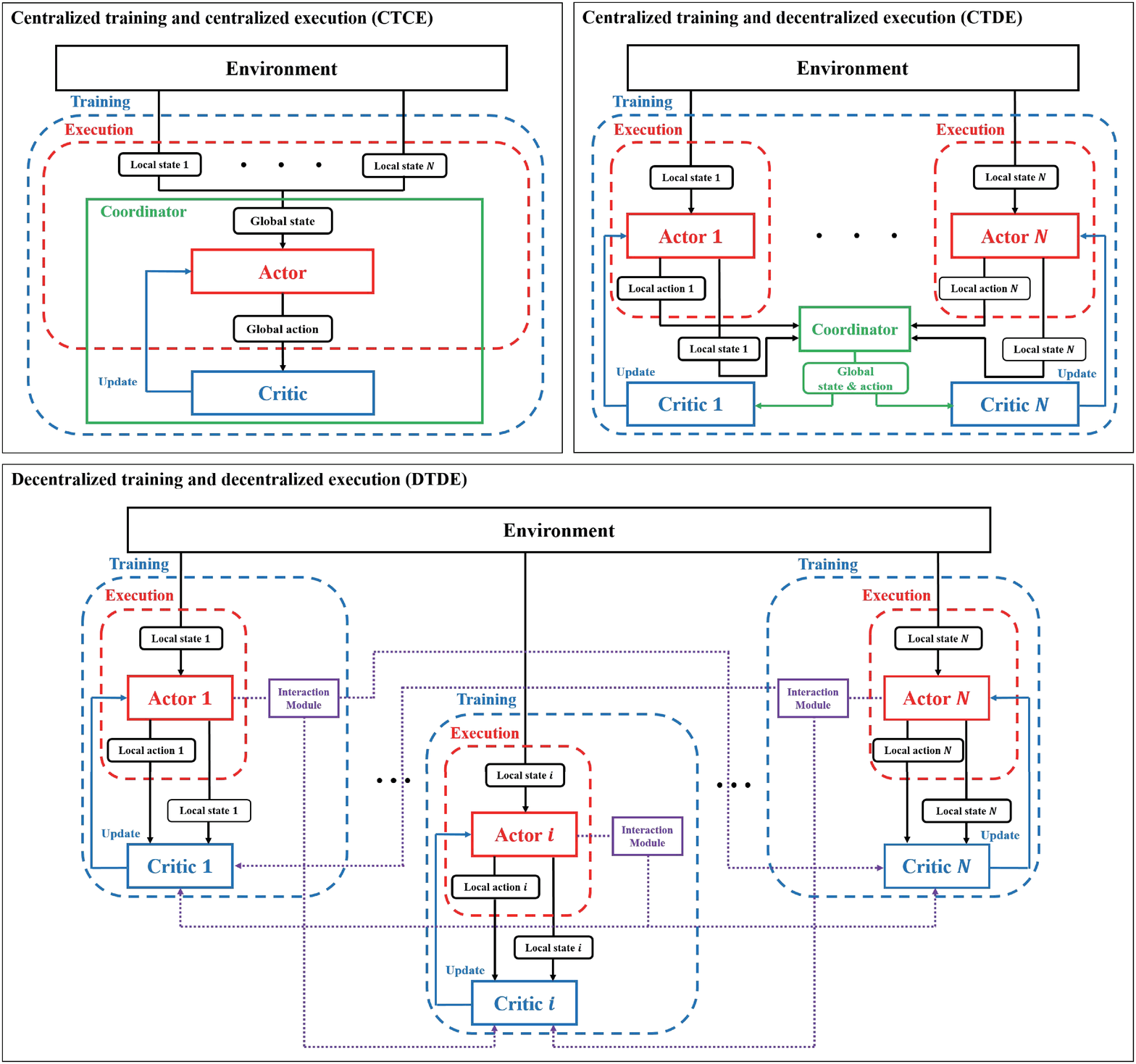}
\end{center}
\caption{Various types of the multi-agent DRL framework. Both the CTCE and CTDE rely on global network information collected from all agents to train a complete set of actor-critic pairs jointly. Such a centralized coordination is avoided in the proposed DTDE structure by means of appropriate interaction modules installed at individual agents. A simple message exchange among agents helps decentralized training of each actor-critic pair in parallel.}
\label{figure:ddpg}
\end{figure*}

\subsection{Learning-Based Design}
DRL approaches have been recently investigated to develop online decision policies of UAV-enabled MEC systems \cite{QLiu:TVT20,LZang:ACC21,LWang:TMC21, LWang:TCCN21, HPeng:JSAC21,AGao:TVT21}. The energy minimization problem of a single UAV server was tackled using the deep Q-network (DQN) framework \cite{QLiu:TVT20,LZang:ACC21}. Hovering positions of the UAV and device scheduling policies were optimized in \cite{QLiu:TVT20}. The method in \cite{LZang:ACC21} determined the task offloading ratio and UAV moving directions such that the energy consumption is minimized. Optimization variables are defined as an action to be determined by a DQN agent that is modelled as a deep neural network (DNN). This UAV agent takes an action based on a state of the environment such as device locations and channel attenuation. For a given network state, the UAV agent chooses the best action that returns the largest Q-value. Such a selection mechanism needs a discrete action space with finite alphabet. Therefore, the DQN methods \cite{QLiu:TVT20,LZang:ACC21} can only identify a particular UAV location among finite numbers of position candidates, which brings a severe performance degradation. 

The deep deterministic policy gradient (DDPG) framework can address DRL problems with continuous-valued actions. It is thus promising for the realistic design of UAV trajectories and task offloading ratios. The operations of the UAV agent are separated into two distinct components, namely, actor and critic, which are constructed by DNNs. The actor straightforwardly generates continuous-valued actions, and its Q-value is estimated by the critic. Such an actor-critic architecture improves the generalization ability to continuous action spaces. The DDPG techniques have been employed in multi-UAV MEC systems \cite{LWang:TMC21,LWang:TCCN21,HPeng:JSAC21,AGao:TVT21}. Major features of existing schemes are summarized in Table \ref{tab:tab1}. The method in \cite{LWang:TMC21} determined the UAV trajectory to minimize the energy consumption of IDs. After optimizing the UAV movement policy via the DDPG, resource management and device scheduling were additionally optimized through conventional matching algorithms while fixing the trajectory. The performance of this separate optimization approach is limited since it cannot capture the coupled nature of trajectory, computation, and communication variables. Therefore, this method restricts the degree of freedom for UAV agents, blocking their potential in designing multiple MEC features jointly.

The UAV trajectory optimization has also been tackled in \cite{LWang:TCCN21,AGao:TVT21}. A simple greedy policy was adopted in \cite{LWang:TCCN21} for the task offloading, and computing resources were straightforwardly assigned according to the offloaded task volume. A game-theoretic task offloading optimization preceded the trajectory decision of trained actors \cite{AGao:TVT21}. The management policy of spectrum, computation, and caching resources has been proposed in \cite{HPeng:JSAC21} without trajectory optimization. The joint optimization of trajectory, communication, and computation strategies of UAV servers has not been studied adequately yet. The actor structure of these conventional studies incurs challenges in handling the Cartesian product of continuous-valued action sets. The entire action space consists of infinitely many combinations of the communication, computation, and mobility variables. This enormous action space is intractable via the naive DNN construction. Therefore, a novel actor architecture is required for the joint optimization of UAV-aided MEC networks.

\begin{figure*}
\begin{center}
\includegraphics[width=\linewidth]{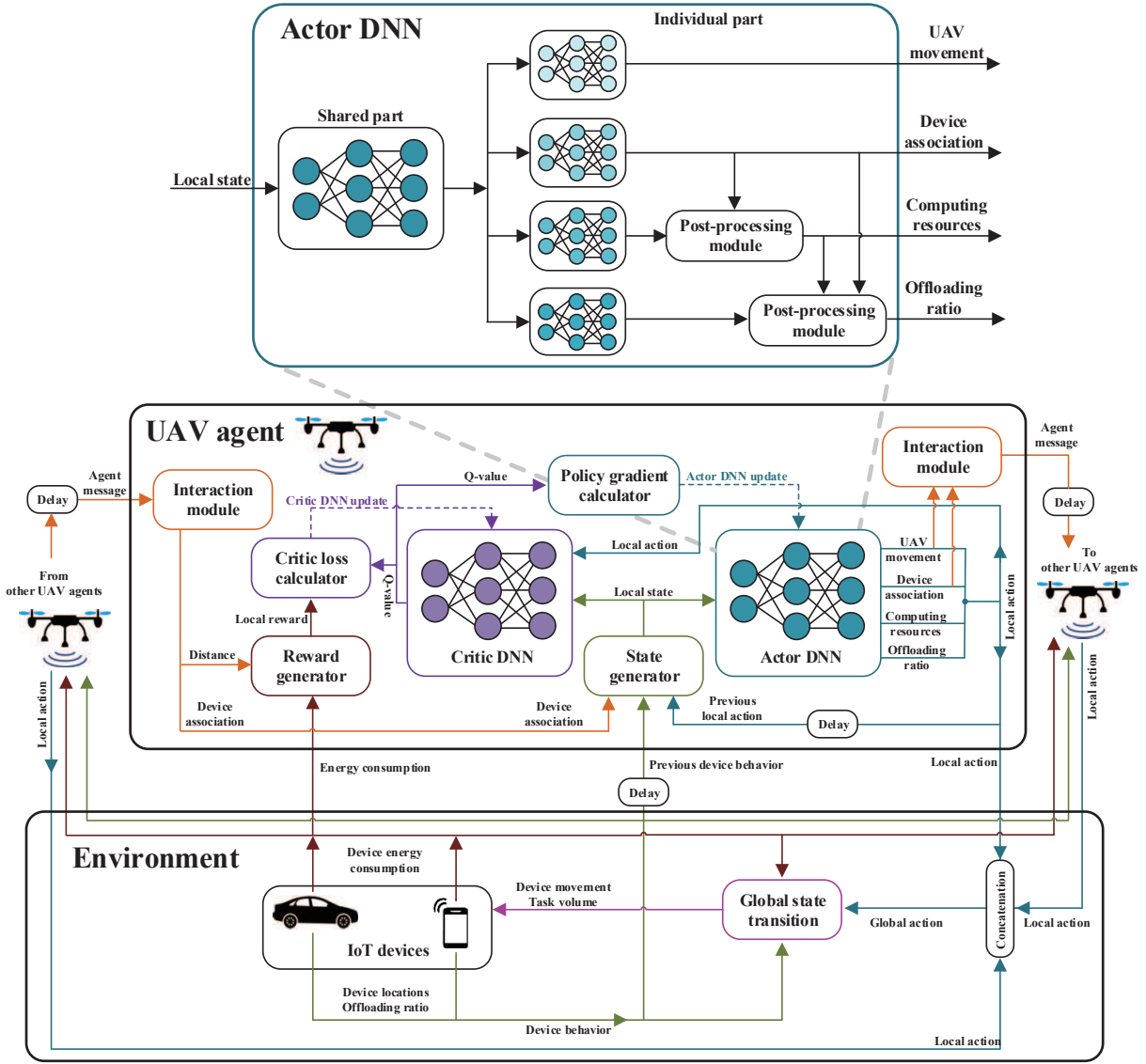}
\end{center}
\caption{Proposed DTDE multi-agent DDPG framework. UAV agents individually collect a local state consisting of their own status, device behaviors, and agent messages shared from other agents. A multi-branch actor architecture facilitates a simultaneous decision of heterogeneous action variables. These collectively form a global action incurring state transitions of the environment. The return of local actions is measured via a local reward which accounts for the energy consumption of associated devices and the penalty for the UAV collision. This is leveraged for the decentralized optimization of an individual actor-critic pair.}
\label{figure:actor}
\end{figure*}

\subsection{Multi-Agent Learning Methods}
The use of multiple UAV servers motivates the development of multi-agent learning systems. Figure~\ref{figure:ddpg} illustrates
three main multi-agent DRL paradigms based on the actor-critic framework. The scheme in \cite{LWang:TMC21} considered a \textit{centralized training and centralized execution (CTCE)} strategy, where a single actor-critic pair is installed at a central coordinator that controls a group of UAVs simultaneously. The actor-critic DNNs work only with global network statistics which are the collections of local states and actions of individual UAVs. To realize this, the coordinator keeps collecting inputs and outputs of UAVs and IDs both in the training and execution steps, incurring centralized DNN calculations. Hence, this architecture is not viable in IoT systems that are typically not included in the network coverage and operate without the coordination of cellular base stations.

A decentralized control mechanism of multiple UAVs has been studied in \cite{LWang:TCCN21,HPeng:JSAC21,AGao:TVT21} by means of the multi-agent DDPG technology, in particular, under the \textit{centralized training and decentralized execution (CTDE)} framework shown in Fig.~\ref{figure:ddpg}. Each UAV server is implemented by an individual agent consisting of its own actor and critic. Multiple UAV agents independently interact with the environment and obtain their local states. As a result, online decision-making calculations of trained actor DNNs can be made possible in a decentralized manner. For a communication-efficient design, the CTDE can be performed with a global data buffer, which is composed of a complete set of local states and local actions of all UAVs. Such a centralized data collection mechanism invokes joint training processes of actor-critic DNNs. Without the central coordinator, the CTDE cannot facilitate an online fine-tuning of separate UAVs in the presence of fluctuating network dynamics.

To eliminate the dependency of centralized calculations, a \textit{decentralized training and decentralized execution (DTDE)} 
framework is necessary for practical MEC systems. In this framework, UAV agents are trained and executed
without the help of the central coordinator. To realize the DTDE policy, agents are carefully designed to use locally available statistics, e.g., local state and local action. By doing so, individual agents can be managed with their dedicated local data buffers. However, the coupled nature of agent behaviors and environment normally requests global network information for learning the optimal policies at individual agents.
To overcome such an insufficient information problem, the DTDE framework employs a proper interaction mechanism among agents that enables message exchanges relevant to decentralized operations. These agent messages should not be the entire set of local statistics which may cause high communication overhead. Thus, agent messages need to be designed to encapsulate minimal sufficient statistics for successful decentralized processing.

\section{Decentralized Management of UAV Servers}\label{sec:proposed MA scheme}

\subsection{Problem Description}
For UAV-aided MEC networks in this article, tasks of all IDs are required to be completed within one system frame. The behaviors of UAV servers should be updated whenever the IDs change their locations. To cope with this time-varying nature, each frame is divided into multiple time slots as in \cite{SJeong:TVT18,QLiu:TVT20,LZang:ACC21,LWang:TMC21,LWang:TCCN21,AGao:TVT21,HPeng:JSAC21} so that the device mobility can be tracked at the beginning of each time slot. UAV servers fly within a region defined in the 3D Euclidean space. Their mobility is subject to the maximum speed limit. To avoid collisions, the minimum distance constraints are added for all pairs of UAVs. Ground IDs and UAVs exchange data packets through air-to-ground channels. By moving UAVs close to IDs, the wireless access distance can be minimized, thereby decreasing the communication latency in air-to-ground links. However, line-of-sight components are dominant at high altitudes, but at the same time, the access distance from IDs gets larger. Such a channel setup involves a nontrivial tradeoff in designing 3D UAV trajectories.

Devices receive computational tasks with randomized data volumes. The partial offloading strategy is adopted for handling separable tasks \cite{JPark:21,Mao:17}. Mobile applications employ the data-partition model to generate independent task bits \cite{Mao:17}. One portion of the task bitstreams is executed locally, whereas the remaining part is offloaded to associated UAVs. Distinct frequency resources are occupied for multiple access of IDs. The UAV assigns dedicated virtual machines (VMs) for each computation request with heterogeneous sizes. To handle this asymmetry, different levels of CPU frequencies can be allocated to individual tasks~\cite{Mao:17}. The VM-based architecture provides an independent system to each ID, thereby paralleling the task execution processes. The total CPU frequency of all VMs is subject to the maximum capacity of a physical machine. The resultant data packets after the computation are then sent back to the corresponding devices.

Practical MEC networks need to minimize the energy consumption of energy-constrained IDs, such as transmission power and local computation cost. At the same time, the overall MEC process should be completed in a timely manner. This imposes a latency constraint on the total completion time composing of the uplink/downlink transmission delays through wireless access channels, which are related to the upload/download of offloaded tasks, as well as the computation time. Such coupled missions request a challenging joint optimization formulation of heterogeneous actions, including the UAV trajectory, device-UAV association, computing resource allocation, and task offloading policy.

\subsection{Joint Learning Strategy of Heterogeneous UAV Actions}
For the joint optimization, an actor should yield different types of MEC actions simultaneously. This needs a higher level of exploration ability to learn desirable policies in uncountable action spaces. Figure~\ref{figure:actor} illustrates the proposed DTDE multi-agent DRL framework along with the multi-branch actor structure. The proposed actor DNN consists of two neural computations: shared part and individual parts. The shared part is first applied to an input local state including the positions of IDs, computing resource allocation and the task offloading policy at the previous time instance, and remaining data of devices to process. Four individual parts follow the shared part, where an individual part is dedicated to the action on each optimization variable. Such a multi-branch structure partitions an entire action space comprising all decision variables into four individual action sets, thereby reducing populations of action candidates. The coupled nature among device association, computation resources, and data offloading variables requests a nested design of individual parts such that an action output of each individual part can affect those of the subsequent parts. To capture this, post-processing modules are embedded at the end of the third and fourth individual parts to mix coupled actions. These guarantee the feasibility of output actions and prevent agents from exploring an infeasible region in the action space.

The first individual part determines the UAV trajectory under the maximum speed constraint. The second individual part decides the device association status, which can be regarded as preference of the UAV towards IDs for computation offloading. A disjoint matching constraint is imposed so that each device is supported by one UAV. The device association is further leveraged for the third individual part to allocate computing resources. By doing so, UAV agents assign CPU computations only to scheduled devices. The resulting output represents the computation capacity, i.e., the CPU frequency, for each task to be offloaded. Finally, this is combined into the fourth individual part which makes the task offloading status. Based on the computing capacity obtained by the third individual part, UAV agents determine the amount of data offloaded from IDs. The offloading ratio is, in general, proportional to the device association preference. To inject this intuition, the output of the second individual part is combined with that of the third one to make the final offloading decision.

\subsection{Decentralized Learning Strategy of Cooperative UAVs}

The proposed actor DNN is installed at individual UAVs, which builds a multi-agent system shown in Fig.~\ref{figure:actor}. Separate UAVs request a novel decentralized learning strategy both in training and execution steps without a central coordinator. This DTDE setup invokes restrictions on agent operations such that they can only leverage locally observable statistics. This lacks global network information and makes individual agents focus only on their own status, thereby ending up an egoistic policy for local agents. A key enabler for this challenge is an interaction mechanism where UAV agents cooperatively adjust their behaviors through message exchanges. This is a distinct feature of the proposed DTDE architecture in comparison with the conventional CTDE framework \cite{LWang:TCCN21,HPeng:JSAC21,AGao:TVT21} which assumes an ideal central coordination in the training step.

The DTDE needs a novel decomposition learning formalism which splits the entire learning problem into several local learning formulations dedicated to individual UAV agents. Each of local states comprises local observations measured from the surrounding nature, i.e., IDs and neighboring UAVs. A local observation of a particular UAV includes
device behavior information shared from IoT nodes such as their previous positions and remaining task volume. The previous location and offloading decision of a UAV are also included in the local observation.

The local observations are not sufficient for correct decisions since they can only form partial information of the global network state. This issue can be resolved by UAV interactions. An interaction module embedded at each UAV generates agent messages that are exchanged with others through air-to-air control channels. The agent message of a particular UAV aggregates its own knowledge that is relevant for appropriate action calculations at others.
The agent message consists of UAV locations and device association solution. Such low-dimensional messages can be shared through point-to-point UAV communication links in parallel with the task execution procedure. Upon the interaction, each UAV constructs its local state with agent messages and local observations. This local state is processed by an actor DNN to calculate a local action. The local actions of all UAVs collectively form a global action. Along with current device behaviors, the global action determines the status of the environment in the future. This incurs the transition of the global network state such as remaining task volumes and device locations in the future.

For the decentralized control, UAV agents evaluate their local actions using individual reward values, which are referred to as local rewards. The proposed local reward aggregates the energy consumption of ground IDs. It is additionally regulated by relative distance from other UAVs not to crash with them. These regularization terms can either be passively measured or collected via agent interactions. Each UAV agent is then optimized by using its corresponding local reward. 

Training UAV agents is also realized in a decentralized manner. Among various DRL frameworks, the DDPG technique is employed in this article. Each UAV agent first collects a local state from the environment and nearby UAVs. This information is passed to the actor DNN yielding local actions. Individual critic DNNs supervise decentralized optimization of associated actor DNNs by estimating the Q-value of local actions and local states. This is the major feature of the proposed DTDE in comparison with the CTDE method that requires global network information. A set of local state, action, reward, and their transitions can be combined into a local data buffer of each UAV to facilitate batch training processes. The estimated Q-value is leveraged to calculate policy gradients for updating the actor DNN. The critic DNN is optimized according to the critic loss so that its Q-value estimates become accurate with training iterations. Consequently, each actor-critic pair is trained in parallel by means of interactions among UAV agents. This DTDE learning strategy enables an online decentralized optimization of the UAV-aided MEC system. 

Once trained, actor DNNs suffice for decentralized decisions. Optimized actors installed at each UAV and take local actions using local information and agent messages. The device association and computing resource actions are leveraged to prepare the VMs with different CPU frequency setups. Also, at the beginning of each time slot, device-specific actions including UAV association and offloading are informed to desired ground IDs through control channels. This establishes air-to-ground offloading links over which IDs send offloaded data packets to scheduled UAVs.

\vspace{5mm}
\section{Performance Evaluation}
\subsection{Implementation Setup}
The advantage of the proposed DTDE strategy is investigated using numerical experiments. Two UAV servers are considered with six IDs randomly deployed over a $100\ \text{m} \times 100\ \text{m}$ square region. They move towards randomized directions with arbitrary speed generated by the first-order Gaussian model \cite{SBatabyal:15Surveytuts}. The duration of one system frame is set to $2$ seconds, and the frame is divided into 10 time slots. Each training episode consists of one frame, whereas the trained agents are tested over four consecutive frames. The frequency division multiple access protocol is adopted for the task exchange where the system bandwidth of $40 \ \text{MHz}$ is evenly assigned to the IDs. The 3D air-to-ground channel model in \cite{AAl:WCL14} is adopted. The maximum computing capacity is $50$ GHz. The volume of computational tasks is uniformly distributed within $[1.2, 12]\ \text{Mbits}$ where the execution of a single bit requires $1550.7$ CPU cycles. The UAV speed is limited by $50 \ \text{m/s}$. The transmission power of IDs and UAVs are respectively given as $0.1\ \text{W}$ and $10\ \text{W}$ \cite{LWang:TCCN21}.

The effectiveness of the proposed joint optimization strategy is examined in comparison with conventional separated design approach \cite{LWang:TMC21}. The DDPG with a naive actor validates the proposed actor DNN architecture. Also, the ideal CTCE and CTDE baselines provide upper bound performance for the proposed DTDE strategy. To see the impact of the agent cooperation, the DTDE method is also realized without any message exchanges among UAVs.

\subsection{Results}

\begin{figure}
\begin{center}
\includegraphics[width=0.55\linewidth]{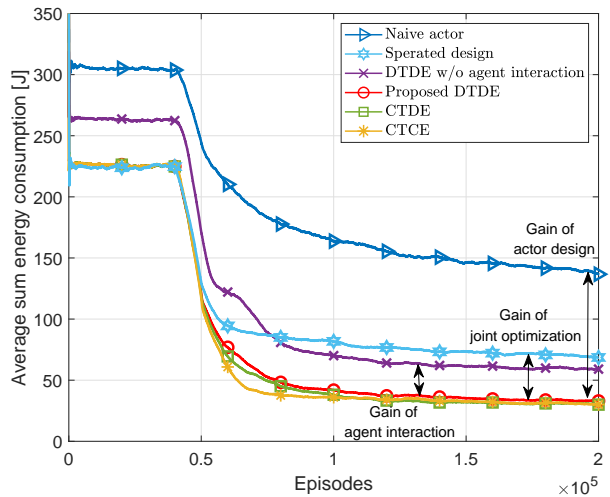}
\end{center}
\caption{Performance comparison of various DDPG techniques. The critic has four fully-connected layers each with 512, 256, 128, and 64 neurons. The actor consists of three fully-connected layers with all 128 neurons for the shared part and two fully-connected layers each with 64 and 32 neurons for the individual parts.}
\label{figure:convergence}
\end{figure}

Figure \ref{figure:convergence} compares various DDPG schemes by assessing the average energy consumption performance with respect to training episodes. Albeit the decentralized training structure, the proposed DTDE quickly converges to ideal centralized baselines, i.e., the CTCE and CTDE methods, validating the effectiveness of the proposed decentralized training mechanism. This implies that practical MEC networks can be efficiently managed without the central coordinator by means of properly designed UAV cooperation and DTDE learning policies. It can be seen that the DTDE without the agent cooperation doubles the device energy consumption. Thus, the impact of UAV interaction plays an important role in the proposed DTDE strategy. The gain of the joint optimization strategy can be examined by comparing with the separated design \cite{LWang:TMC21}. About $50$ J of the energy consumption can be saved using the proposed joint optimization policy. The naive actor baseline provides the worst performance although it determines all the optimization variables jointly. This indicates that the proposed multi-branch actor architecture is essential to identify heterogeneous actions simultaneously.

\begin{figure*}[!htp]
\begin{center}
\includegraphics[width=\linewidth]{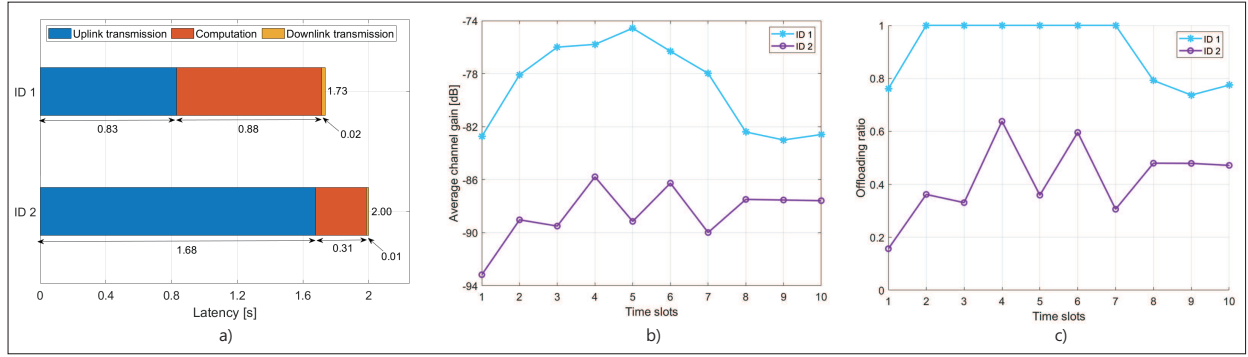}
\end{center}
\caption{Optimized behaviors of the proposed method a) task completion time of two IDs with different task volume b) average channel gain between IDs and their associated UAVs c) optimized task offloading ratio.}
\label{figure:performance1}
\end{figure*}

Figure \ref{figure:performance1} shows the optimized actions of two IDs with asymmetric task sizes. ID 1 and ID 2 need to process tasks of 8.2 and 10.5 Mbits, respectively, within 10 time slots of 2 seconds. The downlink transmission latency is negligible since the UAVs normally have sufficient transmit power compared to the IDs \cite{LWang:TMC21,LWang:TCCN21}. Thus, the overall latency is dominated by the uplink transmission delay and the computation delay. The plot shows that albeit the unbalanced task volumes, both IDs achieve similar completion time satisfying the target latency constraint. A large data volume of ID 1 can be addressed by positioning the associated UAV closely, which improves the average channel gain of ID 1. At the same time, ID 1 offloads the most of its task bits. Nevertheless, thanks to a proper UAV positioning, the uplink transmission latency of ID 1 is fairly reduced compared to ID 2. However, this incurs a high computation delay at the UAV to execute a large volume of the offloaded task. In contrast, the UAV does not care much about ID 2 which has a smaller task volume. The channel gain of ID 2 is not that improved, and thus the latency is dominated by the uplink task transmission. Instead, ID 2 handles about a half of its task locally to fulfill the latency constraint. Consequently, it is concluded that the proposed approach manages a large task volume by dispatching the UAV servers closely, and at the same time, increasing the offloading ratio.

\section{Concluding Remarks and Future Research Opportunities}
\label{sec:conclusions}
This article has presented research directions for DRL-based UAV-aided MEC networks. A DTDE learning strategy has been proposed for the decentralized optimization and management of multiple UAV servers. The computation structure of UAV agents is carefully designed so that it tackles the joint optimization of trajectory, task offloading, and resource allocation policies. A group of these UAVs cooperatively adjust their behaviors individually. Numerical results have demonstrated the proposed mechanism.

Current multi-agent systems are dedicated to a particular configuration with fixed populations of network entities. This lacks the scalability of the system size. Such an issue mainly comes from fixed computation rules of agents. A novel neural architecture, which is flexible to input and output dimensions, is necessary for versatile multi-agent systems. A graph neural network technique could be a promising solution for handling arbitrary network configurations. 

The proposed solution relies on man-made agent messages. Such a handcraft design may become a burden for examining relevant statistics manually. Therefore, a self-organizing interaction policy is essential to realize autonomous DTDE systems. One possible solution is to design agent messages using dedicated DNN units, which have been revealed as an efficient approach in unsupervised learning applications \cite{HLee:JSAC19}. Such a DNN-oriented message is designed only for the desired agents, thereby preserving the private information of users. 

Existing works have been confined to the physical layer design. However, a cross-layer optimization approach is essential to further improve the MEC performance. UAVs serving a number of IDs may have insufficient computing power, thereby increasing the computation delay. To address this challenge, UAVs may additionally offload their received tasks to others with sufficient CPU capacity. This requires a proper task routing protocol that can be investigated as future actions.

\bibliographystyle{IEEEtran}
\begingroup
\vspace*{-1mm}
\bibliography{AZREF}
\endgroup

\end{document}